\documentclass[a4paper,11pt]{article}
\usepackage{jheppub} 
\usepackage[table,xcdraw]{xcolor}   
\usepackage{subfloat}
\usepackage{amsmath,amssymb,amsthm, graphicx, mathrsfs, pgf,tikz,marvosym,fontawesome,slashed, mathtools,array,bbm,yfonts, adjustbox, adforn, fourier-orns, relsize,stmaryrd,simplewick,youngtab,ytableau,youngtab, epigraph, url, bclogo,arydshln,multirow,dsfont, lipsum, tensor,feynmf, wrapfig, enumitem,verbatim}
\usepackage[caption=false]{subfig}
\usepackage{physics}

\numberwithin{equation}{section}

\renewcommand{\d}{\textrm{d}}

\newcommand{\A}{\mathrm{A}}
\newcommand{\B}{\mathrm{B}}
\newcommand{\C}{\mathrm{C}}
\newcommand{\D}{\mathrm{D}}
\newcommand{\E}{\mathrm{E}}
\newcommand{\F}{\mathrm{F}}
\newcommand{\G}{\mathrm{G}}
\newcommand{\HH}{\mathrm{H}}
\newcommand{\I}{\mathrm{I}}
\newcommand{\J}{\mathrm{J}}
\newcommand{\M}{\mathrm{M}}
\newcommand{\PP}{\mathrm{P}}
\newcommand{\R}{R}
\newcommand{\GS}{\Phi}

\newcommand{\ph}{\varphi}

\newcommand{\dph}{\delta \tensor{\ph}{}}

\newcommand{\dpsi}{\tensor{\psi}{}}

\newcommand{\dA}{\tensor{A}{}}
\newcommand{\dB}{\tensor{B}{}}
\newcommand{\dE}{\tensor{E}{}}
\newcommand{\Ppsi}{\Pi_{\psi}}
\newcommand{\Pdph}{\Pi_{\dph}}

\newcommand{\Pf}{\Pi_{f}}
\newcommand{\PE}{\Pi_{E}}
\newcommand{\PR}{\Pi_{R}}
\newcommand{\PPh}{\Pi_{\Phi}}

\newcommand{\Hh}{\mathcal{H}}
\newcommand{\kh}{k}

\newcommand{\phd}{\Dot{\varphi}}

\newcommand{\dphd}{\delta \tensor{\Dot{\ph}}{}}

\newcommand{\dpsid}{\tensor{\Dot{\psi}}{}}

\newcommand{\dEd}{\tensor{\Dot{E}}{}}

\title{Euclidean wormholes stability analysis revisited}

\author[a]{Donald Marolf}
\author[b]{Bruno Missoni}
\affiliation[a]{Department of Physics, University of California, Santa Barbara, CA 93106}
\affiliation[b]{SISSA, Via Bonomea 265, 34136 Trieste, Italy}
\emailAdd{marolf@ucsb.edu, bmissoni@sissa.it}

\abstract{Previous studies of linearized stability of asymptotically flat Euclidean axion wormholes found that symmetric modes suffered from divergences.  We show that such divergences were an artifact of a particular way of solving the constraints, and that a full treatment leads to finite actions for such modes.  The modes must thus be included in a stability analysis.  However, since the action for these modes turns out to be positive, this turns out not to affect previous statements about stability of axion wormholes. We also introduce a technique that allows us to show this positivity at a pseudo-analytic level that avoids heavy numerics.  Our techniques should be useful to future studies of stabilities of other wormholes as well.}

\begin{document}
\maketitle
\flushbottom
\section{Introduction}
Wormholes have caught the interest of a large number of researchers in theoretical physics ever since the early days of Euclidean semi-classical gravity \cite{Coleman:1988tj,Giddings:1987cg}. Axion wormholes represent the model which has been studied in great depth, due to their natural embedding in string theory \cite{Giddings:1987cg,Giddings:1989bq} (though see also e.g \cite{Marolf:2021kjc} for other top-down analyses.). Axions are pseudoscalars which have a shift symmetry, which implies that they can be described in the dual two-form formulation with a three-form field strength. In Euclidean signature this leads to a negative kinetic term in the scalar description, which is required to hold the wormhole throat open. In this simplistic setup, wormholes are supported by the axion-charge flux and they connect two flat spaces. However, more general solutions have also been studied, such as those including dilatons \cite{Arkani-Hamed:2007cpn,Bergshoeff:2004pg} (also with dilatons having a mass \cite{Andriolo:2022rxc,Jonas:2023ipa}) and those with non-zero cosmological constant \cite{Gutperle:2002km,Aguilar-Gutierrez:2023ril}. Being a particular example of gravitational instantons, wormholes should affect the vacuum significantly and in fact have many phenomenological applications in particle physics and cosmology (nicely summarized in \cite{Hebecker:2018ofv}). They also lead to many puzzles, the most prominent one being the factorization puzzle in holography \cite{Maldacena:2004rf,Arkani-Hamed:2007cpn}. 

The puzzles can be resolved if the wormholes are declared to be  irrelevant saddles in the gravitational path integral. This requires a classification of saddles a la Coleman: the Euclidean path integral is performed in the saddle-point approximation
\begin{equation}\label{path int generic}
    \int \mathcal{D}g\mathcal{D}\phi \ e^{-S_E[g,\phi]}\approx \sum_{\text{saddles}} \frac{1}{\sqrt{\det \hat{M}}}e^{-S_E[g_0,\phi_0]},
\end{equation}
where $g_0,\phi_0$ represent the metric and matter fields evaluated at the saddle-point, and $\hat{M}$ is the operator of quadratic fluctuations\footnote{See e.g.  \cite{Marolf:2022ntb,Liu:2023jvm} for discussion of ambiguities in the definition of $\hat M$ and their possible effects.}. Coleman \cite{Coleman:1987rm,Coleman:1985rnk} argued that the number of negative modes determines the meaning of the instanton. Namely, if there are no negative modes in the spectrum of $\hat{M}$ the instanton is a meaningful contribution to the path integral (e.g. it lifts the vacuum degeneracy as double-well instantons in quantum mechanics \cite{Coleman:1985rnk}). If there is a single negative mode, it describes a tunnelling effect\footnote{A notable example being the false vacuum decay in QFT \cite{Coleman:1977py,Callan:1977pt}, with the effects of gravitation included in \cite{Coleman:1980aw}.}. Multiple negative modes imply that the saddle in question is not an actual minima of the action, it does not represent a physically meaningful contribution to the path integral and it is "unstable".

While this kind of "axion wormhole stability analysis" has a long and controversial history \cite{Rubakov:1996cn,Alonso:2017avz,Hertog:2018kbz}, recent works \cite{Jonas:2023qle,Loges:2022nuw,Hertog:2024nys} using a variety of approaches\footnote{Ref. \cite{Loges:2022nuw} used the two-form formulation of axion while \cite{Jonas:2023qle} used the pseudoscalar formulation to study axion-dilaton wormholes in the homogeneous sector of dilaton fluctuations. The work of \cite{Hertog:2024nys} considered all modes of axion-dilaton wormholes in both formulations of axions.} have all found that asymptotically flat axion wormholes have no negative modes. However, \cite{Loges:2022nuw,Hertog:2024nys} seem to indicate that even eigenfunctions should be excluded from the spectrum of the quadratic operator $\hat{M}$. In fact, their final quadratic action takes the following form
\begin{equation}\label{quad sing general}
    \delta^2S_E=\int_{-\infty}^\infty \d r \left[\A \dot{\R}^2+\B\dot{R}\R+\C\R^2\right],
\end{equation}
where $r$ is the Euclidean time, $R$ is the gauge-invariant variable comprised out of axion and metric fluctuations and $\A,\B$ and $\C$ are some functions of the background. It turns out that $B$ and $C$ are singular near $r=0$, in fact $B\sim 1/r$ and $C\sim 1/r^2$. However, when they had rewritten the action in terms of the quadratic operator $\hat{M}$
\begin{equation}\label{sturm liouville general}
    \delta^2S_E=\frac{1}{2}\int_{-\infty}^\infty \d r \R \hat{M}\R + \text{boundary terms},
\end{equation}
they found that singularities are removed. The authors of the aforementioned papers argued that even eigenfunctions can be found in the spectrum of $\hat{M}$, but that they give an infinite contribution when inserted back into  \eqref{quad sing general} and thus should be excluded from the spectrum. However, the absence of divergent contributions in \eqref{sturm liouville general} suggests that it may be useful to treat the singularities in \eqref{quad sing general} with more care. This is also suggested by the fact that similar singularities that arise in a closely related approach to studying wormholes turn out to cancel when treated properly \cite{Liu}.

In this paper we resolve the issue of singularities in two different ways. To explain these resolutions, let us recall that 
gravitational degrees of freedom will generally lead to constraints. One can then use these constraints to perform the path integral over some degrees of freedom, effectively imposing the constraints. A common method of analyzing the stability of saddles follows \cite{Garriga:1997wz,Gratton:1999ya,Kol:2006ga} in choosing a complex contour of integration that allows this procedure and then integrating the remaining (constrained) degrees of freedom over a real Euclidean contour\footnote{See, however, comments in \cite{Horowitz}.}. We will show that singularities found in \cite{Loges:2022nuw,Hertog:2024nys} were an artifact of the choice of variables used to solve the constraints in \cite{Loges:2022nuw,Hertog:2024nys} and, in particular, that making a different choice leads directly to \eqref{sturm liouville general} without encountering singularities along the way. Since there are no singularities, there is no reason to exclude even eigenfunctions from the spectrum.

When we include the dilaton, our new choice  seems to present difficulties for writing  the resulting action in a gauge-invariant form. In this way, it seems that the choice made in \cite{Hertog:2024nys} and the appearance of singularities may be  inevitable.  We therefore propose a way of treating these singularities by allowing for excursions in the complex $r$-plane, using a contour which is partly an infinitesimal semi-circle of infinitesimal radius $\epsilon$ going around the singularity. The contribution of the singularity to the integral is then accounted for as in the spirit of the famous identity in complex analysis \cite{Weinberg:1995mt}
\begin{equation}
   \lim_{\epsilon\rightarrow0} \frac{1}{r+i\epsilon}=\mathcal{P}\left(\frac{1}{r}\right)-i\pi\delta(r),
\end{equation}
where $\mathcal{P}$ denotes the Cauchy principal value of the integral. We see that the singularities contribute with a $\delta$-function which has an imaginary coefficient.

One could then ask a natural question: what is the physical meaning of the imaginary part introduced by the singularities? To understand this, one could go back to the discussion on instantons by Coleman \cite{Coleman:1985rnk}. The operator $\hat{M}$ which has a single negative mode would introduce an imaginary part due to the square-root in \eqref{path int generic}, essentially giving an imaginary contribution to the ground state energy, i.e. a decay rate. In this sense, we could say that the imaginary parts which would come from the singularities would be a contribution to the decay rate not from the spectrum itself but from the singularities. Luckily, for axion wormholes both with and without the dilaton included, the quadratic action is such that the imaginary parts cancel exactly. In this sense, there are no new contributions to the decay rate and even eigenfunctions should be included in the path-integral, contrary to what was claimed in \cite{Hertog:2024nys} and \cite{Loges:2022nuw}. In what follows, we will only present the key aspects of the perturbative analysis of the wormhole and refer the reader to \cite{Hertog:2024nys} for details.

We begin in section \ref{sec:rev} with a brief review of axion wormholes and the models to be studied.  Section \ref{sec:pert} then describes the theory of linearized fluctuations about the wormhole saddles, focussing on the issues identified in \cite{Loges:2022nuw,Hertog:2024nys} associated with even perturbations and showing explicitly in dilaton-free models that the singularities were an artifact of the particular method used in \cite{Loges:2022nuw,Hertog:2024nys} to solve the constraints.  Section \ref{sec:sing} then shows that a simple method based on complex analysis can be used to analyze such singularities in general, and to show that their contributions in fact fully cancel in the full class of models studied in \cite{Loges:2022nuw,Hertog:2024nys}. The modes must thus be included in the analysis.     We also introduce a pseudo-analytic technique that allows us to show these modes have positive action without heavy use of numerics.  We then close in section \ref{sec:conclusions} with brief conclusions emphasizing future directions.
\section{Euclidean axion wormholes}
We will consider a low-energy effective theory made up of an axion and  the dilaton, its usual partner in string compactifications, coupled to gravity. The scalar-field part of the action takes the usual form of a non-linear $\sigma$-model
\begin{equation}
    S_E\supset \frac{1}{2}\int_\mathcal{M} G_{IJ}(\phi)\ \star \d\phi^I \wedge \d\phi^J.
\end{equation}
where $G_{IJ}$ is the metric on target space. Axions have a shift symmetry $\chi\rightarrow \chi+\alpha$, meaning that they have a dual description in terms of a two-form field $B_{\mu\nu}$, with three-form field strength $F=dB$. Axions can be related to the three-form via the Hodge dual relation 
\begin{equation}
 \star F\sim \pm \d\chi,
\end{equation}
up to a factor which depends on the inverse metric on the target space. Due to the presence of the Hodge-star map it is easy to see that axions have a negative kinetic term in Euclidean signature\footnote{One can think of this as if axions themselves get Wick-rotated $\chi\rightarrow i \chi$ upon Wick-rotation to Euclidean time \cite{VanRiet:2020pcn,Burgess:1989da}.}. Because of this fact, the metric on the target space $G_{IJ}$ is generally of indefinite signature. Moreover, it is negative-definite in the axionic directions and positive-definite in the saxionic directions (i.e. fields which do not have a shift symmetry and cannot be dualized). The full Euclidean action containing gravity, axion and the dilaton written in the dual formalism reads
\begin{equation}
    S_E= \int_\mathcal{M}\left[-\frac{1}{2\kappa_4^2}\star\mathcal{R}+\frac{1}{2}e^{-b\varphi}\star F\wedge F+\frac{1}{2}\star \d\varphi\wedge \d\varphi+\chi \d F\right],
\end{equation}
where $b$ is the dilaton coupling. Here the axion field $\chi$ is introduced to act as a Lagrange multiplier imposing $dF=0$, such that it locally be written as $F=dB$. As a result, in the path integral one must integrate this $\chi$ over a contour parallel to the {\it imaginary} axis (so that it gives a Dirac delta function imposing the desired constraint)\footnote{\label{foot:chi} In fact, taking the 3-form gauge field to be $U(1)$ valued makes the complex $\chi$-plane periodic in the imaginary direction.  One should then only integrate $\chi$ over a fundamental period.}.   This somewhat awkward convention is chosen so that Euclidean saddles with real $F$ are equivalent to Euclidean saddles with real $\chi$.  One can nevertheless choose to integrate out $F$ instead of $\chi$ to obtain the action 
\begin{equation}\label{action in chi}
     S_E= \int_\mathcal{M}\left[-\frac{1}{2\kappa_4^2}\star\mathcal{R}-\frac{1}{2}e^{b\varphi}\star \d\chi\wedge \d\chi+\frac{1}{2}\star \d\varphi\wedge \d\varphi-\d(e^{b\varphi}\chi \star \d\chi) \right],
\end{equation}
where again $\chi$ should be integrated over an imaginary contour.
The boundary term is crucial since it is what bounds the gravitational action from below \cite{Hertog:2024nys} when the constraints are imposed. The boundary conditions one imposes on the axions are also non-trivial due to their shift symmetry \cite{Burgess:1989da}.  The boundary term in \eqref{action in chi} tells us that we should fix  the Noether charge conserved due to the shift symmetry. It is called the axion charge and it is given by
\begin{equation}
    Q=\frac{1}{2\pi^2}\int_\Sigma F.
\end{equation}
In the scalar field formulation this reads
\begin{equation}
    Q=\frac{1}{2\pi^2}\int_\Sigma \d^3x\sqrt{\gamma} \partial^0\chi,
\end{equation}
where $\gamma$ is the metric induced on $\Sigma$ and $\pi_\chi=\partial^0\chi$ is the Euclidean momentum of the axion. Since the wormhole throat is supported by the axion-charge flux, we wish to impose that the axion charge remains fixed at the boundary, which in turn implies Dirichlet boundary conditions on $F$ or Neumann boundary conditions on $\chi$. Here it is important that we consider asymptotical flat geometries.  In the asymptotically AdS context, such boundary conditions would lead to violations of the CFT unitarity bounds and, correspondingly, the bulk Hamiltonian becomes unbounded below \cite{Andrade:2011dg}.  One would thus expect additional Euclidean negative modes. Saxions have no shift symmetry, so we impose Dirichlet boundary conditions on their variations at the boundary.

We will consider spherically-symmetric wormhole solutions with the following metric ansatz \cite{Arkani-Hamed:2007cpn,Bergshoeff:2004pg}
\begin{equation}
    \d s^2=N(r)^2\d r^2+a(r)^2\d\Omega_3^2,
\end{equation}
where $d\Omega_3^2$ is the metric on $S^3$. We have that the field strength is then of the form
\begin{equation}
    F=Q\mathcal{E},
\end{equation}
where $\mathcal{E}$ is the volume form on the unit three-sphere. In target space,  the solutions lie on a geodesic with the velocity
\begin{equation}
    G_{IJ} \frac{\d\phi^I}{\d h}\frac{\d\phi^J}{\d h}=c,
\end{equation}
where $h(r)$ is the radial harmonic satisfying
\begin{equation}
    \frac{\d h}{\d r}=\frac{N}{a^3}.
\end{equation}
In the ansatz above the Einstein equation reads
\begin{equation}
    \left(\frac{\dot{a}}{N}\right)^2=1+\frac{\kappa_4^2 c}{6a^4}.
\end{equation}
Different solutions are characterized by the sign of the geodesic velocity on target space. Wormholes correspond to timelike geodesic ($c<0$) and in conformal gauge ($N=a$) the analytic solution is simple 
\begin{equation} a^2(r)=\sqrt{\frac{\kappa_4^2\lvert c\rvert}{6}}\cosh(2r).
\end{equation}
When only the axion is turned on we have that $c=-Q^2$. In the two-field model containing the axion and the dilaton, explicit solutions for the fields are given by \cite{Bergshoeff:2004fq,Bergshoeff:2004pg}
\begin{equation}\label{two field}
    e^{b\varphi/2}=\frac{\lvert Q\rvert}{\sqrt{\lvert c\rvert}}\cos\left(\frac{b\sqrt{\lvert c\rvert}}{2}h(r)\right),\quad \chi(r)=\frac{2\sqrt{\lvert c \rvert}}{bQ}\tan\left(\frac{b\sqrt{\lvert c\rvert}}{2}h(r)\right).
\end{equation}
Solutions exist precisely when one respects the regularity conditions \cite{Arkani-Hamed:2007cpn,VanRiet:2020pcn}
\begin{equation}
    b^2< \frac{8}{3}\kappa_4^2,
\end{equation}
which can be understood as the range of $b$-values for which the cos function in \eqref{two field} does not change sign.

\label{sec:rev}
\section{Perturbation theory}

As argued in the introduction, to discuss the stability of the wormhole we should perturb the action up to quadratic order in the fluctuations
\begin{equation}
    g_{\mu\nu}\rightarrow g_{\mu\nu}+\delta g_{\mu\nu},\quad F_{\mu\nu\rho}\rightarrow F_{\mu\nu\rho}+\delta F_{\mu\nu\rho}, \quad \chi\rightarrow \chi+\delta \chi,\quad \varphi\rightarrow \varphi+\delta\varphi.
\end{equation}
Since this was extensively studied in \cite{Hertog:2024nys} both in the two-form and in the scalar-field formalism of axions, here we only summarize the key points of this perturbative analysis and start at the intermediate result from which our analysis continues. For further details on the perturbative analysis, we refer the reader to\footnote{In \cite{Hertog:2024nys} and here, the action was expanded up to quadratic order using Mathematica packages xPand, xTensor, xAct and xPert \cite{xPand,xtensor,Brizuela:2008ra}, with the modification to calculate perturbations in Euclidean signature \cite{Maenaut:2020Git}.} \cite{Hertog:2024nys}.

Perturbation theory around the wormhole is essentially a cosmological perturbation theory in Euclidean signature \cite{Gratton:1999ya,Gratton:2000fj}
\begin{equation}
    \begin{split}
        \d s^2&=(g_{\mu\nu}+\delta g_{\mu\nu})\d x^\mu \d x^\nu\\
        &=a^2[(1+2A)\d r^2+2D_iB \d r \d x^i+((1-2\psi)\gamma_{ij}+2D_iD_jE)\d x^i \d x^j],
    \end{split}
\end{equation}
where $\gamma$ and $D_i$ are the induced metric and covariant derivative on the unit three-sphere, respectively. We will work in the dual formulation of the axions, parametrizing the three-form fluctuations as
\begin{equation}
    \delta F=\frac{1}{6}f\mathcal{E}_{ijk}\d x^i \wedge \d x^j \wedge \d x^k+\frac{1}{2}\mathcal{E}_{ijk}D^kw \d r \wedge \d x^i \wedge \d x^j,
\end{equation}
with similar parametrizations being used in \cite{Loges:2022nuw,Jonas:2023qle}.  In particular, we consider the Euclidean path integral defined by integrating over real three-form fields, whereas the corresponding path integral would correspond to integrating the axion $\chi$ along the imaginary axis.  See also footnote \ref{foot:chi}. Fluctuations transform under infinitesimal coordinate transformations, i.e. gauge transformations of scalar-type
\begin{equation}
    x^\mu \rightarrow x^\mu -\xi^\mu, \quad \xi^\mu=(\xi^0,D^i\xi).
\end{equation}
Metric fluctuations transform as
\begin{equation}
    A\rightarrow A+\mathcal{H}\xi^0+\dot{\xi}^0, \quad B\rightarrow B+\xi^0+\dot{\xi}, \quad \psi\rightarrow\psi-\mathcal{H}\xi^0, \quad E\rightarrow E+\xi
\end{equation}
and matter fluctuations transform as
\begin{equation}
    f\rightarrow f+Q\triangle \xi, \quad w\rightarrow w+Q\dot{\xi}, \quad \delta \varphi \rightarrow \delta\varphi+\xi^0\dot{\varphi}.
\end{equation}
Since any meaningful physical statement should be gauge-invariant, it is sufficient to work with gauge-invariant variables which are linear combinations of matter and metric fluctuations.

Since our wormhole solution was spherically symmetric, it appears convenient to write our fluctuations in terms of spherical harmonics on $S^3$, denoted by $Y_n$ and satisfying
\begin{equation}
    \triangle Y_n=-(n^2-1)Y_n,
\end{equation}
where $\triangle$ is the Laplacian on $S^3$. Taking the $Y_n$ to be orthonormal in $L^2(S^3)$ makes them easy to integrate out of the action. However, the eigenvalues of $\triangle$ will remain in the action and, for simplicity of notation, we will not distinguish between $\triangle$ and its eigenvalue $-(n^2-1)$.

When expanded up to quadratic order, the action does not depend on the derivatives $\dot{A}$ and $\dot{B}$. Thus $A$ and $B$ are not dynamical fluctuations and they are Lagrangian multipliers which impose constraints. In order to make these constraints more transparent, it is convenient to work in the Hamiltonian formalism
\begin{equation}
    \delta^2S_E =\int \d r\left[p\dot{q}-H(p,q)-C^i\mathcal{N}_i(p,q)\right],
\end{equation}
where $C_i$ are the Lagrange multipliers and $\mathcal{N}_i$ are the respective constraints. It is from solving these constraints that the singularities in \cite{Hertog:2024nys,Loges:2022nuw} have appeared. We will show in the next section, in the case where there is only axion and no dilaton, that the singularities are not present if one solves the constraint in a different way.

\subsection{Axion wormholes}
We start from the quadratic action written in the Hamiltonian formalism\footnote{Note that the fluctuation $w$ is not present in this action, this is because it has been eliminated in \cite{Hertog:2024nys,Loges:2022nuw} by solving the constraint imposed by the Lagrange multipler $\delta\chi$. The constraint is the closure of the three-form, i.e. $\dot{f}=\triangle w$.} (Eq. (3.23) in \cite{Hertog:2024nys})
\begin{equation}\label{eq:typo}
\begin{split}
\delta^2 S_E =&\int \d r \bigg[
\Ppsi \dpsid + \PE \dEd + \Pf \Dot{f} + \Pdph \dphd  + \frac{1}{4 \left(\triangle + 3 \kh\right) a^2}
\left( \kh \Ppsi^2
- \frac{3}{\triangle} \PE^2
- 2 \Ppsi \PE \right)\\
& + \frac{a^2}{2} e^{b \ph} \triangle \Pf^2
- \frac{1}{2a^2}\Pdph^2
- \frac{1}{2} \phd \Ppsi \dph 
+ a^2 \left(\triangle + 3 \kh \right) \psi^2 
+ \frac{e^{-b\ph}}{2a^2} \left(f + Q (3 \dpsi - \triangle \dE - b\dph) \right)^2 \\
&- \frac{e^{-b\ph}}{4a^2} Q^2 b^2 \dph^2 
+ \frac{1}{2a^2} \left( \frac{3}{2} (\phd)^2 - \triangle \right) \dph^2
- \dB \left( \PE + Q \triangle \Pf \right)\\
&- \dA \left( \Pdph \phd - \Hh \Ppsi + \frac{Q \, e^{-b \ph}}{a^2} \left( Q( \triangle \dE - 3 \dpsi + \frac{1}{2} b \dph ) - f \right)
+ a^2 \left( 3 \Hh \phd \dph + 2 ( \triangle + 3 \kh ) \dpsi \right) \right) \bigg],
\end{split}
\end{equation}
where we have set $\kappa_4^2=1$.  The momenta are given by
\begin{align}
\Pf &= \frac{e^{-b\ph}}{\triangle a^2} \left( Q \triangle \dB - \Dot{f} \right),& \Ppsi &= a^2 \left(3 \phd \dphd - 6 \dA \Hh - 6 \dpsid +  2 \triangle\left( \dEd - \dB \right) \right), \notag\\
\Pdph &= a^2 \left( \dphd - \phd \dA \right),& \PE &= \triangle a^2 \left( - \phd \dphd + 2 \dA \Hh + 2 \dpsid + 2 \kh \left( \dEd - \dB \right) \right),
\end{align}
where $\Hh=\frac{\dot{a}}{a}$ is the conformal Hubble rate given by
\begin{equation}
\label{eq:cHR}
    \mathcal{H}^2=k+\frac{ c}{6a^4},
\end{equation}
and $k$ is the curvature of the three-space, which is kept explicit even though we will only be interested in the case when $k=1$. As usual in the Euclidean Hamiltonian formalism, the momenta are taken to be integrated over a contour parallel to the imaginary axis (even though they are real on real Euclidean saddles).  We remind the reader that the same is true of the Harmonic oscillator (whose Euclidean Hamiltonian is $-p^2/2+x^2$).

In the method used in \cite{Hertog:2024nys}, the two Lagrange multipliers $A,B$ are taken to lead to two constraints\footnote{Since the momenta in the Euclidean Hamiltonian path integral are imaginary, one may object that, in order to obtain delta-functions by integrating over $A,B$, one must choose contours for the $A,B$ integrals that depend on the other fields.  This appears to nevertheless be standard in applications of this method in the literature and, in the end, it amounts to a (necessarily complicated) prescription to attempt to avoid the conformal factor problem.  Since this is not the focus of the current paper, we will simply follow the literature and apply this procedure below.  However, the reader may find it more satisfying to integrate out the momenta first  (and thus, in effect, to work in the covariant Euclidean formalism) so that the momenta are replaced by their real on-shell values before the $A,B$ integrals are performed, though in that case there are various sign issues that arise that are associated with the conformal factor problem.} 
\begin{equation}\label{constraints}
    \begin{split}
        &\dA:\quad \left( \Pdph \phd - \Hh \Ppsi + \frac{Q \, e^{-b \ph}}{a^2} \left( Q( \triangle \dE - 3 \dpsi + \frac{1}{2} b \dph ) - f \right)+ a^2 \left( 3 \Hh \phd \dph + 2 ( \triangle + 3 \kh ) \dpsi \right) \right)=0 \\
&\dB: \quad \left( \PE + Q \triangle \Pf \right)=0.
    \end{split}
\end{equation}
The authors of \cite{Hertog:2024nys,Loges:2022nuw} chose to integrate out $\Ppsi$ in the constraint imposed by $A$ but, since $\Ppsi$ enters the constraints only through the term $\Hh\Ppsi$, doing so  leads to factors of $1/\Hh$. Evaluating this  on the wormhole solution leads to
\begin{equation}
    \frac{1}{\mathcal{H}}=\frac{1}{\tanh (2r)},
\end{equation}
which is singular at the wormhole throat. Having this in mind, we instead now choose to integrate out $\psi$ and $\Pi_E$ in \eqref{eq:typo} using the constraints in \eqref{constraints}. We momentarily restrict to the pure-axion wormhole case by setting $b=0$, $\varphi=\delta\varphi=0$ and $c=-Q^2$. We integrate by parts the term $\Pi_\psi \dot{\psi}$ in \eqref{eq:typo} effectively replacing
\begin{equation}
\Pi_\psi\dot{\psi}\rightarrow-\psi\dot{\Ppsi}    
\end{equation}
and the boundary term vanishes since the fluctuations vanish at the boundary. One then integrates out $\psi$ using the constraints and introduces the gauge-invariant variable 
\begin{equation}
    R=f-Q\triangle E.
\end{equation}
After integrating out $\psi$, the action still contains many terms with $\dot{\Ppsi}$. To cancel them we add the following total derivative 
\begin{equation}
    -\frac{1}{2}\frac{\d}{\d r}\left[\frac{a^2\Hh \Ppsi^2+2Q\Ppsi R}{3Q^2-2a^4(\triangle+3k)}\right].
\end{equation}
One can check that this term vanishes on the boundary when background solutions are inserted (since $a(r)\sim e^r$ at infinity). Combining the results, one obtains 
\begin{equation}
\label{eq:316}
\begin{split}
    \delta^2 S_E=\int \d r &\frac{1}{4(\triangle+3k)(3Q^2a-2a^5(\triangle+3k))^2}\bigg[ 9Q^4 \Ppsi^2k-12Q^3a^2\Ppsi\dot{R}(\triangle+3k)\\
    &-\triangle (\Pi_f)^2(3Q^2-2a^4(\triangle+3k))^3+2Q^2a^4(\triangle+3k)(9\Hh^2\Ppsi^2-12k\Ppsi^2+2R^2(\triangle+3k))\\
    &+4a^8(\triangle+3k)
    ^2(-3\Hh^2\Ppsi^2+3k\Ppsi^2+2R^2(\triangle+3k))+8Qa^6\Ppsi \dot{R}(\triangle+3k)^2\\
    &+2\Pi_f(3Q^2-2a^4(\triangle+3k))^2(Q\triangle\Ppsi+2a^2\dot{R}(\triangle+3k)) \bigg].
    \end{split}
\end{equation}
One then finds that the momentum conjugate to $R$ is
\begin{equation}
    \Pi_R=\Pi_f-\frac{Q\Ppsi}{3Q^2-2a^4(\triangle+3k)}.
\end{equation}
Upon writing the action in terms of these variables, it takes the very simple form
\begin{equation}
    \delta^2 S_E=\int \d r \left[\Pi_R \dot{R}+\mathcal{A} \Pi_R^2+\mathcal{B}  R^2\right],
\end{equation}
where
\begin{equation}
    \mathcal{A} =\frac{\triangle(-3Q^2+2a^4(\triangle+3k))}{4a^2(\triangle+3k)}\quad \mathcal{B} =\frac{a^2(\triangle+3k)(Q^2+2a^4(\triangle+3k))}{(3Q^2-2a^4(\triangle+3k))^2}.
\end{equation}
The fluctuations  $\Pi_\psi$ and $E$ have completely disappeared from the action, and path integrals over these variables factorize as infinite gauge-orbit volumes which can be reabsorbed in the path-integral normalization \cite{Gratton:1999ya}.

The homogeneous mode ($n=1$) needs to be treated separately. In that case $\mathcal{A} =0$, which means that upon integrating out momenta we impose $\dot{R}=0$, implying that the homogeneous mode is non-dynamical.  Using the boundary conditions then yields $R=0$. For $n=2$, the function $\mathcal{\mathcal{A}}$ diverges. This divergence is already evident in \eqref{eq:316} and arises because the coefficient of $\psi$ in the constraint \eqref{constraints} vanishes for $n=2$. The case $n=2$ must therefore also be treated separately. Indeed, this mode is pure gauge \cite{Gratton:1999ya} in the bulk. However, as argued by \cite{Loges:2022nuw} it also leads to a non-zero total derivative. This total derivative is proportional to the $r\rightarrow \pm \infty$ limit of $\Pi_f^2$. The associated boundary term has the usual (negative) sign for a Euclidean term quadratic in momenta, and thus describes a stable saddle with the prescription that the Euclidean momenta are integrated over the imaginary contour as explained below \eqref{eq:cHR}. We may thus set this case aside and henceforth consider only $n> 2$. 
 
 Integrating out the momenta (for modes $n>2$) imposes the  constraint
\begin{equation}
    \dot{R}+2\mathcal{A} \Pi_R=0,
\end{equation}
which leads to the following action
\begin{equation}\label{quad action axion}
    \delta^2S_E=\int \d r \left[-\frac{1}{4\mathcal{A} }\dot{R}^2+\mathcal{B} R^2\right].
\end{equation}
\begin{figure}[t!]
\centering
 \subfloat[$-1/4\mathcal{A}$]{\includegraphics[width=0.4\textwidth]{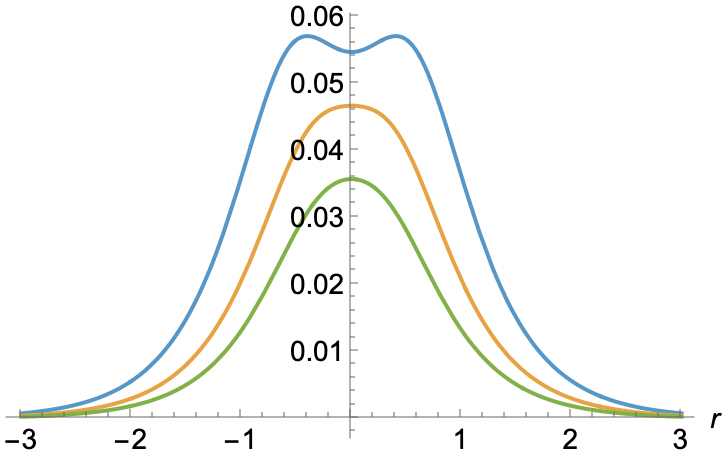}}\hspace{40pt}
 \subfloat[$\mathcal{B}$]{\includegraphics[width=0.4\textwidth]{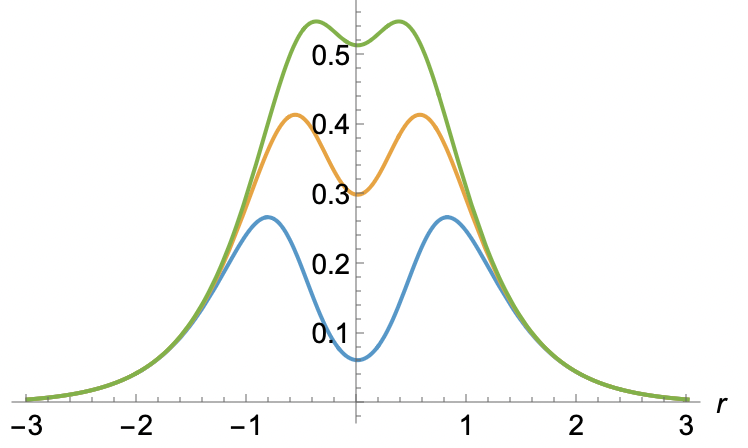}}
\label{axionwhs}
\caption{Functions in \eqref{quad action axion} for $n=3$ (blue), $n=4$ (orange) and $n=5$ (green) ($Q=1$).}
\end{figure}
The same result is obtained in \cite{Loges:2022nuw,Hertog:2024nys} before they performed the rescaling of the fluctuations. The functions multiplying the two squared terms in \eqref{quad action axion} are positive-definite (Figure \ref{axionwhs}); all modes increase the action and all eigenvalues are positive. There is a key difference, however, between our analysis and the ones in \cite{Loges:2022nuw,Hertog:2024nys}: at no point we encountered any singularities in the action. So if one was to define a Sturm-Liouville problem from \eqref{quad action axion} and determine the spectrum, it is clear that one should include both even and odd modes, contrary to what was claimed in \cite{Loges:2022nuw,Hertog:2024nys}.

Let us now bring back the dilaton and look at the constraints \eqref{constraints} with the dilaton included. It turns out that by choosing to integrate out $\psi$ we cannot form a gauge-invariant variable for the dilaton. In \cite{Hertog:2024nys} dilaton forms a gauge-invariant variable with $\psi$
\begin{equation}
    \Phi=\delta\varphi+\frac{\dot{\varphi}}{\mathcal{H}}\psi.
\end{equation}
After  integrating out $\psi$ we have no metric fluctuation which can form a gauge-invariant variable with $\delta\varphi$. However, the momentum $\Pf$ and $\delta\varphi$ can form the following gauge-invariant fluctuation
\begin{equation}
    \Phi_{f}=\delta\varphi-\frac{a^2 \dot{\varphi}}{Q}e^{b\varphi}\Pi_f.
\end{equation}
One needs to remember that we have to impose Dirichlet boundary conditions on the dilaton at $r\rightarrow\pm\infty$. Looking at the asymptotic behaviour of this variable
\begin{equation}
    \Phi_{f}\xrightarrow{r \rightarrow \pm \infty}\delta\varphi+\frac{Q\sin\left(\sqrt{\frac{3}{8}}{b}\pi\right)}{2\sqrt{\lvert c \rvert}}\Pi_f.
\end{equation}
This means that, if we would impose Dirichlet boundary conditions on $\Phi_f$, we would not impose Dirichlet boundary conditions on the dilaton, but rather a mixed boundary condition between the dilaton and the momentum $\Pf$. This is incompatible with the boundary conditions we impose in the path integral, so this cannot be the proper gauge-invariant description.

It is clear that when the dilaton is included, the procedure of writing the action in a consistent gauge-invariant way becomes an increasingly difficult task. It is likely that instead of simply integrating out $\psi$ from the $A$-constraint in \eqref{constraints}, one should integrate out a linear combination
\begin{equation}
    \psi+\alpha ( \beta_1 \Pi_\psi \ + \beta_2\Pi_{\delta\varphi}), 
\end{equation}
for some $\beta_1,\beta_2$ and $\alpha$. The coefficient $\alpha$ should vanish when the dilaton is turned off, so that one recovers  the simple case of axion wormholes where simply $\psi$ is integrated out from the constraint.  Rather than explore this complicated procedure, one could instead choose to live with the singularities arising in \cite{Loges:2022nuw,Hertog:2024nys} and try to resolve them in a different way. Since the singularities are an artifiact of the method of solving the constraints, an appropriate bookkeeping technique should allow one to see that the singularities in fact cancel. As we will see, there is simple way to do this using elementary complex analysis.

\label{sec:pert}
\section{Singularity resolution}
In this section we will resolve the singularities which arise in the quadratic actions of \cite{Hertog:2024nys,Loges:2022nuw}. We will work with axion-dilaton wormholes using the results from \cite{Hertog:2024nys} and the results then immediately hold for the more trivial case of axion wormholes in \cite{Loges:2022nuw}. Our starting point is the quadratic action written in the gauge-invariant formalism (Eq. (3.28) in \cite{Hertog:2024nys}),
\begin{align}\label{gauge inv three-form hamiltonian}
\delta^2 S_E =&\int \d r \Bigg[\PR \Dot{R} + \PPh \Dot{\Phi}
+ \PR \left(
- \frac{e^{-b \ph} Q^2 \triangle}{2 a^4 \Hh (\triangle + 3 \kh)} R
+ \frac{Q \triangle}{4(\triangle+3\kh)} \left(\frac{b e^{-b\ph}Q^2}{a^4 \Hh} + 6 \phd \right) \Phi
\right)\notag\\
&+ \PPh \left(
\frac{Q \triangle \phd}{2a^2 \Hh \left(\triangle + 3 \kh \right)} \PR
- \frac{e^{-b\ph} Q \kh \phd}{2 a^4 \Hh^2 (\triangle + 3 \kh)} R
+ \frac{\phd\left(b e^{-b \ph} Q^2 \kh- 2 a^4 \Hh \triangle  \phd \right)}{4 a^4 \Hh^2 (\triangle + 3 \kh)} \Phi
\right)\notag\\
&+ \frac{\triangle \left(-3Q^2 + 2 e^{b \ph} a^4 (\triangle + 3 \kh) \right)}{4 a^2 (\triangle + 3 \kh)} \PR^2
+ \frac{e^{-2 b \ph} \left(-Q^2 \triangle + e^{b \ph} a^4 (\triangle + 3 \kh) \left(6 \kh + (\phd)^2\right)\right)}{12 a^6 \Hh^2 (\triangle+3\kh)} R^2\notag\\
&+ \frac{e^{-2b\ph} \left( b Q^3 (2 \triangle +3 \kh) - 2 e^{b\ph} Q a^4 \left( - 3 \Hh \triangle \phd + b (\triangle + 3 \kh) (6 \kh + (\phd)^2)  \right) \right)}{12 a^6 \Hh^2 (\triangle + 3 \kh) } R \Phi \notag\\
&+\frac{e^{-2 b \ph}}{48 a^6 \Hh^2 (\triangle + 3 \kh)}\Big( - b^2 Q^4 (2\triangle + 3 \kh) - e^{2b\ph} \triangle a^8 (6 \kh + (\phd)^2) (2 (\triangle+3\kh) + 3 (\phd)^2)\notag\\
&+e^{b \ph} Q^2 a^4 (2(\triangle + 3 \kh) (\triangle + 3 b^2 \kh) - 6 b \triangle \Hh \phd + (3 \triangle + b^2 \triangle+ 3 b^2 \kh)(\phd)^2) \Big) \Phi^2\notag\\
&+\frac{e^{-b\ph} \left(Q^2 (\triangle + 3 \kh) - e^{b \ph} a^4 (6 \kh (\triangle + 3 \kh)+ \triangle (\phd)^2)\right)}{12 a^6 \Hh^2 (\triangle + 3 \kh)} \PPh^2
\Bigg].
\end{align}
where
\begin{equation}
    R=f-Q\triangle E,\quad \PPh=\delta\varphi+\frac{\phd}{\Hh}\psi
\end{equation}
and
\begin{align}\label{actionRphi}
\PR &= \Pf + \frac{Q e^{-b \ph}}{a^2 \Hh}\psi,&
\PPh &= \Pdph - \left(\frac{b e^{-b \ph}}{2a^2 \Hh} + 3 a^2 \phd  \right) \dpsi.
\end{align}
One can see many $1/\Hh$ factors appearing in \eqref{gauge inv three-form hamiltonian}, which lead to a singularity at $r=0$. We proceed without worrying about singularities until the very end when we define the Sturm-Liouville problem. We rewrite the action above schematically in terms of functions $\A_n,\dots,\J_n$
\begin{equation}\label{actionfunctions1}
\begin{split}
    \delta^2S_E=\int \d r &\bigg[\Pi_\R\Dot{\R}+\Pi_\Phi \Dot{\Phi}+\A_n\Pi_\R^2+\B_n\Pi_\R\R+\C_n\R^2+\D_n \Pi_\Phi^2+\E_n\Pi_\Phi \Phi\\
    &+\F_n\Phi^2+\G_n\Pi_\Phi\Pi_\R+\HH_n\Pi_\R\Phi+\I_n \Pi_\Phi \R+\J_n \R \Phi\bigg],
\end{split}
\end{equation}
where by $n$ we explicitly indicate the dependence on the $S^3$-mode. These functions are listed in Appendix \eqref{functions}. In summary, there are three distinct cases to consider \cite{Hertog:2024nys}:
\begin{itemize}
    \item $n=1$: As in the case of axion wormholes, axion fluctuations are not dynamical in the homogeneous sector. However, the dilaton is dynamical and, once the momentum has been integrated out, we find the following quadratic action
\begin{equation}\label{hommodeaction}
    \delta^2 S^{n=1}_E=\int \d r \left[-\frac{1}{4\D_1}\Dot{\GS}^2-\frac{\E_1}{2\D_1}\Dot{\GS}\GS+\left(\F_1-\frac{\E_1^2}{4\D_1}\right)\GS^2\right].
\end{equation}
\item  $n=2$: All the terms in \eqref{gauge inv three-form hamiltonian} are divergent and the mode is pure gauge.
\item $n>2$: Both fluctuations are dynamical and, once the momenta have been integrated out, we obtain the following action \cite{Hertog:2024nys}
\renewcommand{\A}{\mathcal{A}_n}
\renewcommand{\B}{\mathcal{B}_n}
\renewcommand{\C}{\mathcal{C}_n}
\renewcommand{\D}{\mathcal{D}_n}
\renewcommand{\E}{\mathcal{E}_n}
\renewcommand{\F}{\mathcal{F}_n}
\renewcommand{\G}{\mathcal{G}_n}
\renewcommand{\HH}{\mathcal{I}_n}
\renewcommand{\I}{\mathcal{J}_n}
\renewcommand{\J}{\mathcal{K}_n}
    \begin{equation}\label{2action}
    \begin{split}
    \delta^2S_E^{n>2} = \int \d r\bigg[&\A\dot{\R}^2+\B \R\dot{\R}+\C \R^2+\D\dot{\GS}^2+\E\dot{\GS}\GS+\F\GS^2\\
    &+\G\R\GS + \HH \dot{\R}\GS+\I\R\dot{\GS}+\J\dot{\R}\dot{\GS}\bigg].
    \end{split}
\end{equation}
The relations between the functions in \eqref{actionfunctions1} and those in \eqref{2action} are given in Appendix \ref{functions}.
\end{itemize}
The modes we should analyze, i.e. $n=1$ and $n>2$, both exhibit singularities at the wormhole throat. One can easily check that singular behaviours are of the form $1/r$ and $1/r^2$. In order to explain how to deal with these singularities, we make a small digression into the area of complex analysis.
\subsection{Moving to the complex plane}

Our action is defined as an integral over the real line, but we will use the trick of writing the integral in the complex plane in order to understand how to deal with the singularities appearing in the action. First, consider a general integral in the complex plane over a contour $C$
\begin{equation}
    \oint_C \d z \ B(z) f(z),
\end{equation}
where $C$ goes along the real axis from $-R$ to $-\varepsilon$, along a semicircle of radius $\varepsilon$, along the real axis from $\varepsilon$ to $R$ and we close the contour with the semicircle of radius $R$, denoted by $C_R$. We will eventually take the limit of $\varepsilon\rightarrow0 $. The function  $B(z)$ has a pole at $z=0$, i.e. $B(z)\sim b_0/z$ for some constant $b_0$. We will add a small imaginary part $i\eta$ to $z$ in order to place the pole below the real axis. In summary, we get
\begin{equation}
\begin{split}
    &\oint_C \d z \ B(z) f(z)\\
    =& \lim_{\substack{R\rightarrow\infty\\
    \varepsilon,\eta\rightarrow 0}} \bigg[\int_{-R}^{-\varepsilon}\d x \ B(x)f(x)+ib_0\int_
    \pi^0\d\theta \ \varepsilon e^{i\theta}\frac{f(\varepsilon e^{i\theta})}{\varepsilon e^{i\theta}+i\eta}+\int_{\varepsilon}^{R}\d x \ B(x)f(x)+\int_{C_R} \d z B(z) f(z)\bigg]\\
    =&\mathcal{P}\left(\int_{-\infty}^{\infty} \d x B(x) f(x)\right)-i b_0\pi f(0)+\lim_{R\rightarrow\infty}\int_{C_R}\d z\ B(z) f(z).
    \end{split},
\end{equation}
where $\mathcal{P}$ denotes the Cauchy principal value of the integral. Notice that the RHS is independent of $\varepsilon$, so it is also valid in the limit of $\varepsilon\rightarrow 0$. The contributions from the outer semicircle of radius R cancel on both sides and we get that the integral over the real line, when $B(x)$ has a pole at $x=0$, should be replaced by
\begin{equation}
    \int_{-\infty}^{\infty} \d x \ B(x) f(x) \rightarrow \mathcal{P}\left(\int_{-\infty}^{\infty} \d x\ B(x) f(x)\right)-i b_0\pi f(0).
\end{equation}
Now consider a slightly different integral
\begin{equation}
    \oint_C \d z \ C(z) f(z), 
\end{equation}
and in this case near $z=0$ we have $C(z)\sim c_0/z^2$. We use similar manipulations to the ones above to show that
\begin{equation}
\begin{split}
    &\oint_C\d z \  C(z) f(z)\\
    =& \lim_{\substack{R\rightarrow\infty\\
    \varepsilon,\eta\rightarrow 0}} \left[\int_{-R}^{-\varepsilon}\d x \ C(x)f(x)+ic_0\int_
    \pi^0d\theta \ \varepsilon e^{i\theta}\frac{f(\varepsilon e^{i\theta})}{(\varepsilon e^{i\theta}+i\eta)^2}+\int_{\varepsilon}^{R}\d x \ C(x)f(x)+\int_{C_R} \d z C(z) f(z)\right]\\
    =&\mathcal{P}\left(\int_{-\infty}^{\infty} \d x C(x) f(x)\right)-i c_0\pi \dot{f}(0)+\lim_{R\rightarrow\infty}\int_{C_R}\d z\ C(z) f(z),
    \end{split}
\end{equation}
where we have used the fact that
\begin{equation}
\begin{split}
    &ic_0 \lim_{\varepsilon, \eta\rightarrow 0}\int_\pi^0 \d\theta \ \varepsilon e^{i\theta}\frac{f(\varepsilon e^{i\theta})}{(\varepsilon e^{i\theta}+i\eta)^2}\\
    =&-c_0 \lim_{\varepsilon, \eta\rightarrow 0}\int_\pi^0 \d\theta \ \frac{\d}{\d \theta}\left[\frac{1}{\varepsilon e^{i\theta}+i\eta}\right] f(\varepsilon e^{i\theta})\\
    =&-c_0 \lim_{\varepsilon, \eta\rightarrow 0}\left[\frac{1}{\varepsilon e^{i\theta}+i\eta}f(\varepsilon e^{i\theta})\bigg\rvert_\pi^0+c_0\int_\pi^0 \d\theta  \frac{1}{\varepsilon e^{i\theta}+i\eta} \frac{\d}{\d \theta} f(\varepsilon e^{i\theta})\right].
\end{split}
\end{equation}
The first  term vanishes in the limit of $\varepsilon\rightarrow0$. After some chain rule manipulations, the second one equals $-i c_0\pi \dot{f}(0)$, where the dot represents the derivative with respect to $x$. Using the same approach as before, when $C(x)$ is singular on the real line the integral over the real axis should be replaced by
\begin{equation}
    \int_{-\infty}^{\infty} \d x \ C(x) f(x) \rightarrow \mathcal{P}\left(\int_{-\infty}^{\infty} \d x \ C(x) f(x)\right)-i c_0\pi \dot{f}(0).
\end{equation}
We will use these results to understand the singularities in the action for the $n=1$ and $n>2$ modes.
\subsection{Homogeneous sector $n=1$}
As in \cite{Hertog:2024nys}, we schematically write the homogeneous mode action as 
\begin{equation}\label{hommodeaction2}
    \delta^2 S^{n=1}_E=\int \d r \left[\alpha\Dot{\GS}^2+\beta\Dot{\GS}\GS+\gamma\GS^2\right],
\end{equation}
where
\begin{equation}
    \alpha=-\frac{1}{4\D_1},\quad \beta=-\frac{\E_1}{2\D_1},\quad \gamma=\F_1-\frac{\E_1^2}{4\D_1}.
\end{equation}
Inserting the background solutions, one sees that near $r=0$ we have
\begin{equation}
    \beta\sim \frac{\sqrt{\frac{3}{2}} b^2 \sqrt{\lvert c\rvert} }{\left(3 b^2-8\right) r} +\mathcal{O}(r)=\frac{\beta_0}{r}+\mathcal{O}(r), \quad \gamma\sim -\frac{\sqrt{\frac{3}{2}} b^2\sqrt{\lvert c\rvert} }{2 \left(3 b^2-8\right) r^2}+\mathcal{O}(1)=\frac{\gamma_0}{r^2}+\mathcal{O}(1),
\end{equation}
and $\alpha$ is regular near $r=0$. Adopting the prescription from the previous section and integrating the second term by parts we obtain
\begin{equation}\label{afterintegrationhom}
    \delta^2 S_E=\mathcal{P}\left[\int \d r \left(\alpha \Dot{\Phi}^2+\left(\gamma-\frac{\dot{\beta}}{2}\right)\Phi^2\right)\right]+i\pi(\beta_0+2\gamma_0)\Phi(0)\dot{\Phi}(0)+\frac{\beta}{2}\Phi^2\bigg\rvert_\partial.
\end{equation}
\begin{figure}[t!]
\centering
 \subfloat[$\alpha$]{\includegraphics[width=0.4\textwidth]{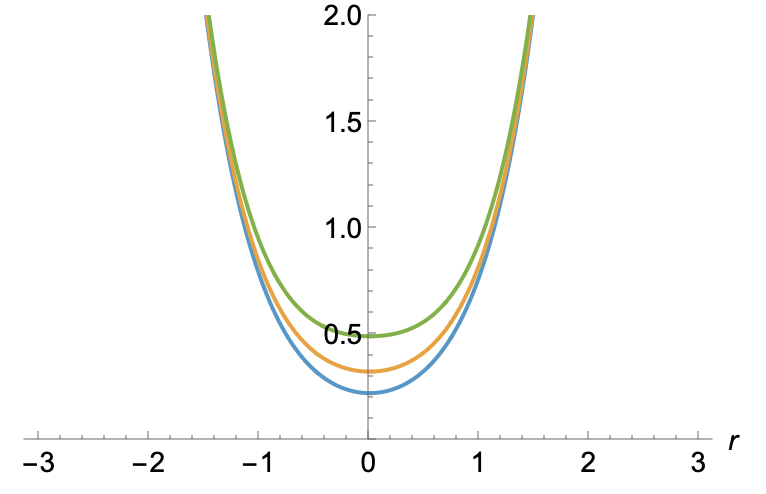}}\hspace{40pt}
 \subfloat[$\gamma-\frac{\dot{\beta}}{2}$]{\includegraphics[width=0.4\textwidth]{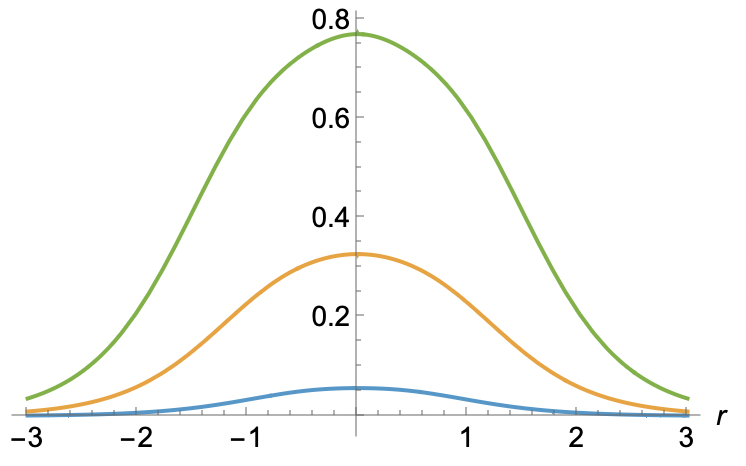}}
\label{hom mode dilaton}
\caption{Functions in \eqref{afterintegrationhom} for $b=0.5$ (blue), $b=1$ (orange) and $b=1.25$ (green). The quadratic action in \eqref{afterintegrationhom} does not depend on $Q$ and we had set $c=-1$.}
\end{figure}
The boundary term vanishes since we impose Dirichlet boundary conditions on the dilaton fluctuations. Additionally, one can easily see that the imaginary contribution in \eqref{afterintegrationhom} also cancels. Moreover, $\gamma-\frac{\dot{\beta}}{2}$ is regular near $r=0$ and there is no need to impose the Cauchy principal value prescription, so we can drop the $\mathcal{P}$-symbol next to the integral. It is worth reemphasizing the importance of the above cancellation of the imaginary terms.  This supports the claim that our detour into the complex plane is merely a useful and transparent book keeping device and that it does not in fact affect the action of our modes.

Since the functions multiplying the quadratic terms are always positive (Figure \ref{hom mode dilaton}), the action always increases and all the modes have positive eigenvalues. However, in contrast to what was claimed in \cite{Hertog:2024nys}, there is no reason to exclude even eigenfunctions using this careful treatment.
\subsection{Inhomogeneous sector $n>2$}
\renewcommand{\A}{\mathcal{A}_n}
\renewcommand{\B}{\mathcal{B}_n}
\renewcommand{\C}{\mathcal{C}_n}
\renewcommand{\D}{\mathcal{D}_n}
\renewcommand{\E}{\mathcal{E}_n}
\renewcommand{\F}{\mathcal{F}_n}
\renewcommand{\G}{\mathcal{G}_n}
\renewcommand{\HH}{\mathcal{I}_n}
\renewcommand{\I}{\mathcal{J}_n}
\renewcommand{\J}{\mathcal{K}_n}

We start from the action \eqref{2action}, it turns out that many functions are singular near $r=0$, namely
\begin{equation}\label{singular inhomogeneous}
\begin{split}
    &\B\sim-\frac{6 \left(\sqrt{6} \sqrt{| c| }\right)}{r \left(Q^2 \left(-9 b^2+4 n^2+20\right)\right)}+\mathcal{O}(r)=\frac{b_0}{r}+\mathcal{O}(r),\\
    &\C\sim-\frac{3 \left(\sqrt{6} \sqrt{| c| }\right)}{r^2 \left(Q^2 \left(9 b^2-4 n^2-20\right)\right)}+\mathcal{O}(1)=\frac{c_0}{r^2}+\mathcal{O}(1),\\
    &\E\sim \frac{3 \sqrt{\frac{3}{2}} b^2 \sqrt{| c| }}{r \left(9 b^2-4 n^2-20\right)}+\mathcal{O}(r)=\frac{e_0}{r}+\mathcal{O}(r),\\
    &\F \sim-\frac{3 \left(\sqrt{\frac{3}{2}} b^2 \sqrt{| c| }\right)}{2 r^2 \left(9 b^2-4 n^2-20\right)}+\mathcal{O}(1)=\frac{f_0}{r^2}+\mathcal{O}(1),\\
    &\G\sim -\frac{3 \left(\sqrt{6} b \sqrt{| c| }\right)}{r^2 \left(Q \left(-9 b^2+4 n^2+20\right)\right)}+\mathcal{O}(1)=\frac{g_0}{r^2}+\mathcal{O}(1),\\
    &\HH\sim-\frac{3 \left(\sqrt{6} b \sqrt{| c| }\right)}{r \left(Q \left(9 b^2-4 n^2-20\right)\right)}+\mathcal{O}(r)=\frac{i_0}{r}+\mathcal{O}(r),\\
    &\I\sim -\frac{3 \left(\sqrt{6} b \sqrt{| c| }\right)}{r \left(Q \left(9 b^2-4 n^2-20\right)\right)}+\mathcal{O}(r)=\frac{j_0}{r}+\mathcal{O}(r).
\end{split}
\end{equation}
Applying the same procedure as before and integrating some terms by parts one obtains the following action
 \begin{equation}\label{2action2}
    \begin{split}
    \delta^2S_E^{n>2} = \mathcal{P}\ \bigg \{\int \d r\bigg[&\A\dot{\R}^2+\left(\C -\frac{\dot{\B}}{2}\right)\R^2+\D\dot{\GS}^2+\left(\F -\frac{\dot{\E}}{2}\right)\GS^2\\
    &+\frac{1}{2}\left(\I-\HH\right)\R \dot{\GS}+\frac{1}{2}\left(\HH-\I\right)\dot{R} \GS\\
    &+\frac{1}{2}\left(\G-\dot{\HH}\right)\R\GS +\frac{1}{2}\left(\G-\dot{\I}\right)\R\GS+\J\dot{\R}\dot{\GS}\bigg]\bigg \}\\
    &-i\pi \big[(b_0+2c_0)\R(0)\dot{\R}(0)+(e_0+2f_0)\GS(0)\dot{\GS}(0)\\
    &+(g_0+j_0)\R(0)\dot{\GS}(0)+(g_0+i_0)\GS(0)\dot{\R}(0)\big]\\
    &+\text{boundary terms}
    \end{split}
\end{equation}
where the boundary terms are expicitly given by
\begin{equation}
\label{eq:BT17}
\bigg \{\frac{\B}{2}\R^2+\frac{\E}{2}\GS^2+\frac{1}{2}(\HH+\I)\R \GS\bigg \} \bigg \rvert_\partial.
\end{equation}
The terms \eqref{eq:BT17} vanish when our boundary conditions are imposed. From the coefficients we defined in \eqref{singular inhomogeneous}, it is straightforward to see that the imaginary parts in $\eqref{2action2}$ again cancel out. Moreover, since all the combinations of functions are now regular, we do not need to restrict the integral to the Cauchy principal value and the $\mathcal{P}$ can be dropped. Again, we see that even eigenfunctions should be included after a careful treatment of the singularities.

One can now show that inhomogeneous fluctuations increase the action. While in \cite{Hertog:2024nys} this was done numerically, for most values of the parameters will be be able to show this here pseudo-analytically by using a simple theorem from linear algebra. We rewrite \eqref{2action2} as a product on a vector space
\begin{equation}
    \delta^2S_E =\int \d r\ v^T\mathcal{M} v,
\end{equation}
where 
\begin{equation}
   v= \begin{pmatrix}
        \dot{R}\\
        \dot{\GS}\\
        \R\\
        \GS
    \end{pmatrix}, \quad \mathcal{M}=\begin{pmatrix}
        \A &  \frac{\J}{2} & 0 & \frac{1}{4}(\HH-\I)\\
        \frac{\J}{2} & \D & \frac{1}{4}(\I-\HH) & 0\\
        0 & \frac{1}{4}(\I-\HH) & \C-\frac{1}{2}\dot{\B} & \frac{1}{2}\left(\G-\frac{1}{2}(\dot{\HH}+\dot{\I})\right)\\
        \frac{1}{4}(\HH-\I) & 0 & \frac{1}{2}\left(\G-\frac{1}{2}(\dot{\HH}+\dot{\I})\right)& \F-\frac{1}{2} \dot{\E}
    \end{pmatrix}.
\end{equation}

Note that the matrix $\mathcal{M}$ is not analogous to the quadratic operator $\hat{M}$ discussed extensively before, since there is no Sturm-Liouville problem involved. If $\mathcal{M}$ is positive-definite, then the action is always increased and this implies that the spectrum of $\hat{M}$ contains only positive eigenvalues. To see that this is the case, we apply the Sylvester criterion \cite{Gilbert}: a symmetric matrix is positive-definite iff all of its principal minors (i.e. determinants of principal submatrices) and $\det \mathcal{M}$ itself are positive. Since we have a 4-by-4 matrix this translates into four conditions
\begin{equation}\label{determinant conditions}
    \begin{split}
        \mathcal{M}_1\equiv&\A>0,\\
        \mathcal{M}_2\equiv&\A\D-\frac{\J^2}{4}>0,\\
        \mathcal{M}_3\equiv& \A\D\left(\C-\frac{1}{2}\dot{\B}\right)-\frac{1}{16}(\I-\HH)^2-\frac{1}{4}\J^2\left(\C-\frac{1}{2}\dot{\B}\right)>0,\\
        \mathcal{M}_4\equiv&\frac{1}{64}\bigg[-16\left(\C-\frac{1}{2}\dot{\B}\right)\left(\frac{1}{4}\D(\I-\HH)^2+\J^2\left(\F-\frac{\dot{\E}}{2}\right)\right)\\
        &+\left(-\frac{1}{2}(\HH-\I)^2+\J(\dot{\HH}+\dot{\I}-2\G)\right)^2\\
        &+4\A\bigg(16\left(\C-\frac{1}{2}\dot{B}\right)\D\left(\F-\frac{1}{2}\dot{\E}\right)-\left(\F-\frac{1}{2}\dot{\E}\right)(\I-\HH)^2\\
        &-\D\left(\dot{\HH}+\dot{\I}-2\G\right)^2\bigg
        )\bigg]>0.
    \end{split}
\end{equation}
\begin{figure}[t!]
\centering
 \subfloat[$\mathcal{M}_1$]{\includegraphics[width=0.4\textwidth]{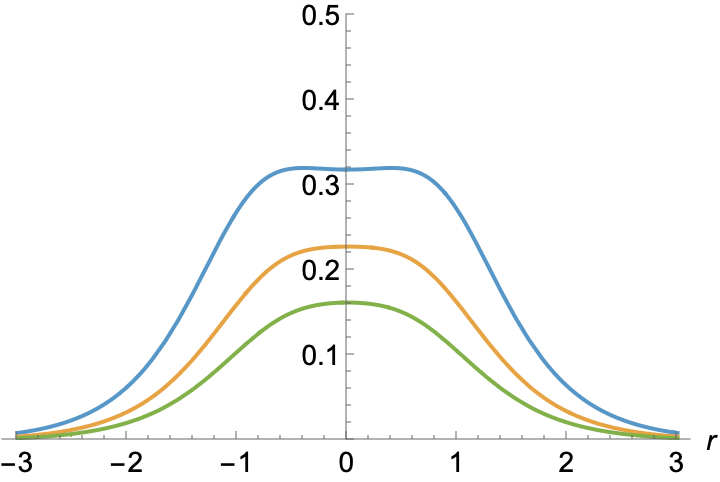}}\hspace{40pt}
 \subfloat[$\mathcal{M}_2$]{\includegraphics[width=0.4\textwidth]{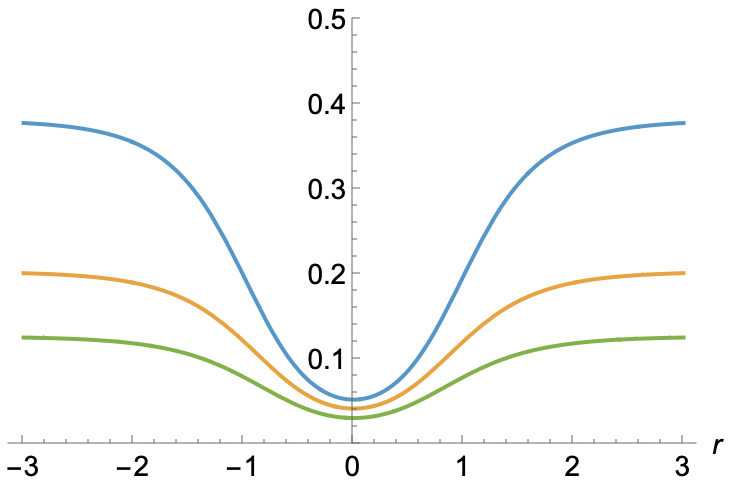}}\\
 \subfloat[$\mathcal{M}_3$]{\includegraphics[width=0.4\textwidth]{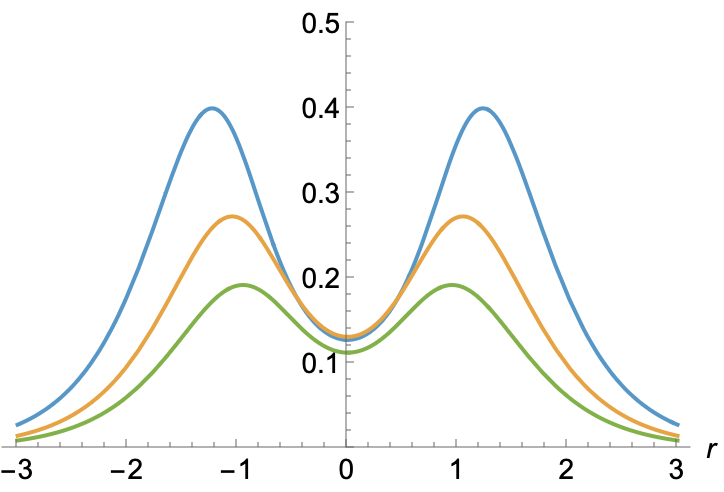}}\hspace{40pt}
 \subfloat[$\mathcal{M}_4$]{\includegraphics[width=0.4\textwidth]{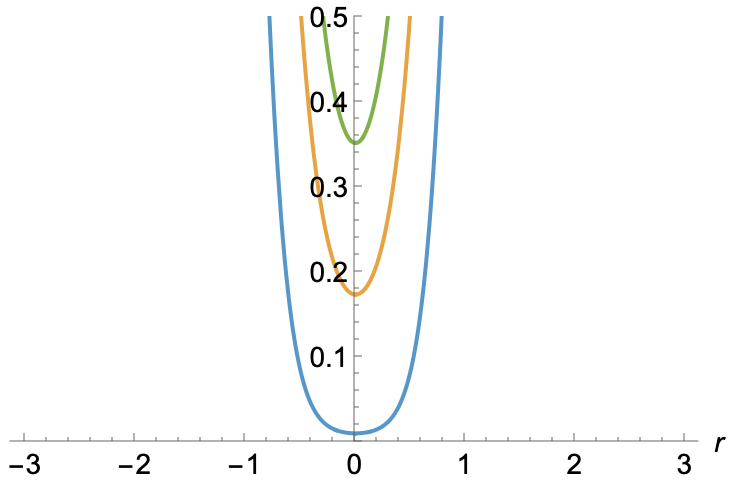}}
\caption{Principal minors from \eqref{determinant conditions} for $n=3$ (blue), $n=4$ (orange) and $n=5$ (green). Other parameters have been set to $c=-1$, $Q=0.5$ and $b=1$.}
\label{determinants}
\end{figure}

\begin{figure}[t!]
\centering
 \subfloat[$\mathcal{M}_4$]{\includegraphics[width=0.5\textwidth]{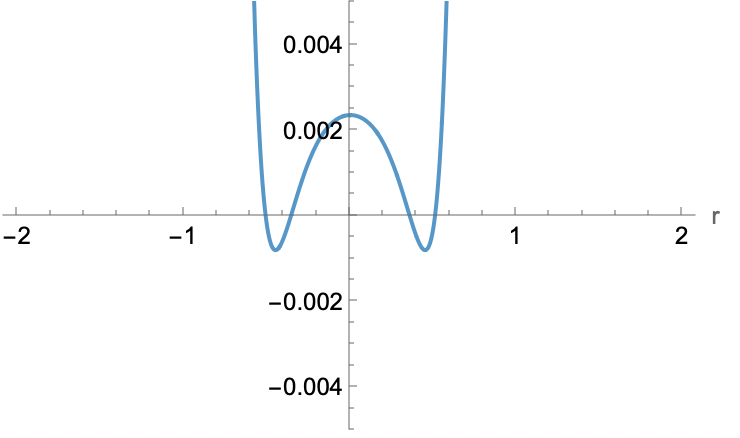}}
\caption{The determinant $\mathcal{M}_4$ plotted for $b=1.45$, $Q=0.5$, $c=-1$ and $n=3$.}
\label{determinant m4}
\end{figure}

These quantities have been plotted in Figure \ref{determinants} and we can see that the matrix $\mathcal{M}$ is positive-definite and all the eigenvalues are positive. However, for the $n=3$ mode, around the value $b\sim 1.45$ the determinant $\mathcal{M}_4$ is negative over a certain domain (Figure \ref{determinant m4}), so the result remains inconclusive. To this end, we also report a few numerical results in the Appendix \ref{numerical eigenvalues} to show that the eigenvalues are indeed positive in the range not captured by this analytic proof.

\label{sec:sing}
\section{Conclusion}
\label{sec:conclusions}

The results reported above place  the stability of axion wormholes on firmer footing by including the eigenvalues of even eigenfunctions in the spectrum. We hope that this way of treating the singularities found in \cite{Loges:2022nuw,Hertog:2024nys} will also be useful in  more general  contexts. Indeed, since the fundamental source of the diverging factor $1/{\cal H}$ is that fact that gradients vanish at the wormhole throat, similar issues will arise in studying perturbations about any $Z_2$-symmetric wormhole.

We have also  described a pseudo-analytic method for analysing the positive-definiteness of the quadratic action. This may be especially useful in cases where the analytic solution in conformal gauge is not always available, such as axion wormholes with cosmological constant or massive dilaton, or for wormholes like those constructed in \cite{Marolf:2021kjc} that are supported by more complicated matter fields. The stability of wormholes with massive dilaton in the homogeneous sector was discussed in \cite{Jonas:2023qle} and it is natural to try to use Sylvester criterion to discuss the stability in the inhomogeneous sector. 
While de Sitter axion wormholes have been argued to be perturbatively stable \cite{Aguilar-Gutierrez:2023ril}, the AdS case has not been directly explored.  Furthermore, the AdS case remains important to investigate both due to its intrinsic interest and because the fact that requires Dirichlet boundary conditions for the pseudoscalar axion in contrast to the Neumann condition used here and in \cite{Hertog:2024nys}.  The importance of the Dirichlet boundary condition in AdS can be seen e.g. from the fact that Neumann boundary conditions would lead to a scalar operator in the dual CFT with scaling dimension zero and which would therefore violate unitarity bounds.  In particular, Neumann boundary conditions for the axion lead to ghosts in the Lorentzian bulk theory, and thus to Euclidean negative modes  \cite{Andrade:2011dg}.

One can also try to discuss the stability of wormholes for more general $\sigma$-models as well as for wormholes like those constructed in \cite{Marolf:2021kjc}. Since  $\sigma$-model wormhole solutions are characterized by geodesics in target space, one can adapt the coordinate system to lie along the geodesic \cite{Hertog:2024nys}. We believe that by using the Sylvester criterion one could deduce restrictions on possible geometries of target spaces which lead to stable wormhole solutions.

As a final remark, recent work on axion wormholes seems to suggest  that these are indeed relevant saddle points in the path integral. If this result persists in AdS with Dirichlet boundary conditions on the axion, the  puzzles these wormholes   bring  to AdS/CFT are not resolved by perturbative instabilities and should thus remain active areas of study. Moreover, since axion wormholes contain no negative modes, it remains important to understand what is their precise meaning in the path integral.
\section*{Acknowledgements}
We thank Thomas Hertog, Simon Maenaut and Thomas Van Riet for helpful discussions. BM would especially like to thank Simon Maenaut for the permission to use his Mathematica code for the perturbative analysis of Section 3. DM especially thanks Xiaoyi Liu for conversations that led to the start of this project.  DM was supported in part by NSF grant PHY-2408110 and by funds from the University of California.
\appendix
\section{Functions in Section 4.}\label{functions}
The functions in \eqref{actionfunctions1} are given by
\renewcommand{\A}{\mathrm{A}_n}
\renewcommand{\B}{\mathrm{B}_n}
\renewcommand{\C}{\mathrm{C}_n}
\renewcommand{\D}{\mathrm{D}_n}
\renewcommand{\E}{\mathrm{E}_n}
\renewcommand{\F}{\mathrm{F}_n}
\renewcommand{\G}{\mathrm{G}_n}
\renewcommand{\HH}{\mathrm{H}_n}
\renewcommand{\I}{\mathrm{I}_n}
\renewcommand{\J}{\mathrm{J}_n}
\begin{equation}
\begin{split}
        &\A=\frac{\triangle \left(-3Q^2 + 2 e^{b \ph} a^4 (\triangle + 3 \kh) \right)}{4 a^2 (\triangle + 3 \kh)}, \quad \B=- \frac{e^{-b \ph} Q^2 \triangle}{2 a^4 \Hh (\triangle + 3 \kh)},\\
        &\C=\frac{e^{-2 b \ph} \left(-Q^2 \triangle + e^{b \ph} a^4 (\triangle + 3 \kh) \left(6 \kh + (\phd)^2\right)\right)}{12 a^6 \Hh^2 (\triangle+3\kh)},\\
        &\D=\frac{e^{-b\ph} \left(Q^2 (\triangle + 3 \kh) - e^{b \ph} a^4 (6 \kh (\triangle + 3 \kh)+ \triangle (\phd)^2)\right)}{12 a^6 \Hh^2 (\triangle + 3 \kh)},\\
        &\E=\frac{\phd\left(b e^{-b \ph} Q^2 \kh- 2 a^4 \Hh \triangle  \phd \right)}{4 a^4 \Hh^2 (\triangle + 3 \kh)},\\
        &\F=\frac{e^{-2 b \ph}}{48 a^6 \Hh^2 (\triangle + 3 \kh)}\Big( - b^2 Q^4 (2\triangle + 3 \kh) - e^{2b\ph} \triangle a^8 (6 \kh + (\phd)^2) (2 (\triangle+3\kh) + 3 (\phd)^2)\\
        &\quad\quad +e^{b \ph} Q^2 a^4 (2(\triangle + 3 \kh) (\triangle + 3 b^2 \kh) - 6 b \triangle \Hh \phd + (3 \triangle + b^2 \triangle+ 3 b^2 \kh)(\phd)^2) \Big), \\
        &\G= \frac{Q \triangle \phd}{2a^2 \Hh \left(\triangle + 3 \kh \right)}, \quad \HH=\frac{Q \triangle}{4(\triangle+3\kh)} \left(\frac{b e^{-b\ph}Q^2}{a^4 \Hh} + 6 \phd \right),  \quad \I=- \frac{e^{-b\ph} Q \kh \phd}{2 a^4 \Hh^2 (\triangle + 3 \kh)},\\
        &\J= \frac{e^{-2b\ph} \left( b Q^3 (2 \triangle +3 \kh) - 2 e^{b\ph} Q a^4 \left( - 3 \Hh \triangle \phd + b (\triangle + 3 \kh) (6 \kh + (\phd)^2)  \right) \right)}{12 a^6 \Hh^2 (\triangle + 3 \kh) }.
\end{split}
\end{equation}
Their relations to the ones in \eqref{2action} are given by
\begin{equation}
\begin{split}
    &\mathcal{A}_n=-\frac{\D}{4\A\D-\G^2},\quad
    \mathcal{B}_n=\frac{\G\I-2\B\D}{4\A\D-\G^2} \quad \mathcal{C}_n=\frac{-\B^2\D+4\A\C\D-\C\G^2+\B\G\I-\A\I^2}{4\A\D-\G^2},\\
    &\mathcal{D}_n=-\frac{\A}{4\A\D-\G^2}, \quad \mathcal{E}_n=\frac{\G\HH-2\A\E}{4\A\D-\G^2}, \quad \mathcal{F}_n=\frac{-\E^2\A+4\A\F\D-\F\G^2+\E\G\HH-\D\HH^2}{4\A\D-\G^2},\\
    &\mathcal{G}_n=\frac{\B(\E\G-2\D\HH)+\G\HH\I-\G^2\J+\A(4\D\J-2\E\I)}{4\A\D-\G^2}, \\
    &\mathcal{I}_n=\frac{\E\G-2\D\HH}{4\A\D-\G^2},\quad \mathcal{J}_n=\frac{\B\G-2\A\I}{4\A\D-\G^2}, \quad \mathcal{K}_n=\frac{\G}{4\A\D-\G^2}.
\end{split} 
\end{equation}
\section{Numerical eigenvalues for the ground state eigenfunctions}\label{numerical eigenvalues}
\renewcommand{\A}{\mathcal{A}_n}
\renewcommand{\B}{\mathcal{B}_n}
\renewcommand{\C}{\mathcal{C}_n}
\renewcommand{\D}{\mathcal{D}_n}
\renewcommand{\E}{\mathcal{E}_n}
\renewcommand{\F}{\mathcal{F}_n}
\renewcommand{\G}{\mathcal{G}_n}
\renewcommand{\HH}{\mathcal{I}_n}
\renewcommand{\I}{\mathcal{J}_n}
\renewcommand{\J}{\mathcal{K}_n}
We follow closely the method explained in \cite{Hertog:2024nys} to determine the eigenvalues of the quadratic operator in \eqref{2action2}. It appears convenient to rescale the axion fluctuation 
\begin{equation}
    R(r)=\sqrt{\cosh{2r}}\ \PP(r)
\end{equation}
which means  $\PP$ should decay at least as $\PP(r)\sim e^{-\lvert r\rvert}$ at the boundaries to be consistent with the boundary conditions.
Finally, writing \eqref{2action} as 
\begin{equation}
    \delta^2S_E=\int \d r \begin{pmatrix}
    \PP & \GS
    \end{pmatrix}\hat{\M}\begin{pmatrix}
    \PP \\
    \GS
    \end{pmatrix}+\mathrm{boundary}\ \mathrm{terms},
\end{equation}
gives us the matrix Sturm-Liouville problem
\begin{equation}
    \hat{\M}\begin{pmatrix}
        \PP \\
        \GS
    \end{pmatrix}=\lambda
    \begin{pmatrix}
        \PP \\
        \GS
    \end{pmatrix},
\end{equation}
where $\hat{\M}$ is a self-adjoint matrix differential operator whose matrix entries are\cite{Hertog:2024nys}
\begin{equation}\label{sturmliouvillematrixelements}
\begin{split}
    \M_{11}&=-\frac{\d}{\d r}\left (\A\frac{\d}{\d r}\right)+\C-\frac{1}{2}\dot{\B}, \\ \M_{12}&=-\frac{1}{2}\J\frac{\d^2}{\d r^2}+\frac{1}{2}(\I-\HH-\dot{\J})\frac{\d}{\d  r}+\frac{1}{2}\left(\G-\dot{\HH}\right),\\
    \M_{21}&=-\frac{1}{2}\J\frac{\d^2}{\d r^2}+\frac{1}{2}(\HH-\I-\dot{\J})\frac{\d}{\d r}+\frac{1}{2}\left(\G-\dot{\I}\right), \\
    \M_{22}&=-\frac{\d}{\d r}\left (\D\frac{\d}{\d r}\right)+\F-\frac{1}{2}\dot{\E},
\end{split}
\end{equation}
and the functions $\mathcal{A}_n,\dots,\mathcal{K}_n$ are not the same as in \eqref{2action} but they are changed due to rescaling
\begin{equation}\label{functions after rescaling}
\begin{split}
   &\A \ \rightarrow \ \A \cosh{2r}, \quad \B \ \rightarrow \ \B \cosh{2r}+2\A \sinh{2r},\\
   &\C \ \rightarrow\ \C \cosh{2r}+\B \sinh{2r}+\A\sinh{2r}\tanh{2r}\\,
   &\G \ \rightarrow \ \sqrt{\cosh{2r}}\ \G +\frac{\sinh{2r}}{\sqrt{\cosh{2r}}}\ \HH,\\
   &\HH\ \rightarrow \ \sqrt{\cosh{2r}}\HH, \quad \I \ \rightarrow \ \sqrt{\cosh{2r}}\ \I +\frac{\sinh{2r}}{\sqrt{\cosh{2r}}}\ \J, \quad \J \ \rightarrow \sqrt{\cosh{2r}}\ \J.
\end{split}
\end{equation}
 \begin{table}[t!]
\begin{center}
\centering
\begin{tabular}{||c | c c  | c c|  c c ||} 
 \hline
  & \multicolumn{2}{|c|}{$b=0.5$} & \multicolumn{2}{|c|}{$b=1$}  & \multicolumn{2}{|c|}{$b=1.5$} \\ 
  \hline\hline
$S^3$ Mode  & $\lambda$& $\PP_0/\GS_0$ & $\lambda$& $\PP_0/\GS_0$ &   $\lambda$ & $\PP_0/\GS_0$  \\ [0.5ex] 
 \hline
 $n=3$ & 0.83399  & 1.85909   & 0.577774  & 1.11246 & 0.367605  &0.849919   \\
 $n=4$  & 1.75748 & 2.08380    & 1.34563 & 1.29989  &0.973307  & 1.04194\\
 $n=5$ &  2.53844&2.65972  &  2.11149  & 1.59748  & 1.66659  & 1.27199 \\
 $n=6$ & 3.12698&  3.53721& 2.7747  &  2.01777 & 2.34006 & 1.56070\\  
 \hline 
 \hline
\end{tabular}
\caption{Eigenvalues and the values of the ratio $\PP_0/\GS_0$ for which the asymptotic boundary conditions hold, for four different $S^3$ modes and three values of $b$, with $Q=0.5$ and $c=-1$.}
\label{eigenvalues:axdil}
\end{center}
\end{table}
One can easily see that this does not change the cancellation of singularities in \eqref{2action2}.
The boundary terms in the action are explicitly given by
\begin{equation}
\bigg \{\A \PP \Dot{\PP}+\frac{\B}{2}\PP^2+\D\GS \Dot{\GS}+\frac{\E}{2}\GS^2+\frac{\J}{2}(\PP\Dot{\GS}+\Dot{\PP}\GS)+\frac{1}{2}(\HH+\I)\PP \GS\bigg \} \bigg \rvert_\partial,
\end{equation}
and the only non-trivial boundary-term is $\mathcal{D}_n\Phi\dot{\Phi}$, which requires $\Phi$ to decay faster than $e^{-\lvert r \rvert}$ at the boundaries. The eigenvalues for the ground state and for different values of $b$ are given in Table \ref{eigenvalues:axdil}. We obtain these using a generalization of the shooting method for two eigenfunctions, explained extensively in the Appendix B of \cite{Hertog:2024nys}. The crux is that since this is a 2D Matrix Sturm-Liouville problem, we need two shooting parameters, which in our case are the eigenvalue $\lambda$ and $\PP_0/\GS_0$, i.e. the ratio of the values of eigenfunctions at $r=0$.

\bibliographystyle{JHEP}
\bibliography{biblio.bib}

\providecommand{\href}[2]{#2}\begingroup\raggedright\begin{thebibliography}{10}

\bibitem{Coleman:1988tj}
S.R.~Coleman, \emph{{Why There Is Nothing Rather Than Something: A Theory of the Cosmological Constant}}, \href{https://doi.org/10.1016/0550-3213(88)90097-1}{\emph{Nucl. Phys. B} {\bfseries 310} (1988) 643}.

\bibitem{Giddings:1987cg}
S.B.~Giddings and A.~Strominger, \emph{{Axion Induced Topology Change in Quantum Gravity and String Theory}}, \href{https://doi.org/10.1016/0550-3213(88)90446-4}{\emph{Nucl. Phys. B} {\bfseries 306} (1988) 890}.

\bibitem{Giddings:1989bq}
S.B.~Giddings and A.~Strominger, \emph{{STRING WORMHOLES}}, \href{https://doi.org/10.1016/0370-2693(89)91651-1}{\emph{Phys. Lett. B} {\bfseries 230} (1989) 46}.

\bibitem{Marolf:2021kjc}
D.~Marolf and J.E.~Santos, \emph{{AdS Euclidean wormholes}}, \href{https://doi.org/10.1088/1361-6382/ac2cb7}{\emph{Class. Quant. Grav.} {\bfseries 38} (2021) 224002} [\href{https://arxiv.org/abs/2101.08875}{{\ttfamily 2101.08875}}].

\bibitem{Arkani-Hamed:2007cpn}
N.~Arkani-Hamed, J.~Orgera and J.~Polchinski, \emph{{Euclidean wormholes in string theory}}, \href{https://doi.org/10.1088/1126-6708/2007/12/018}{\emph{JHEP} {\bfseries 12} (2007) 018} [\href{https://arxiv.org/abs/0705.2768}{{\ttfamily 0705.2768}}].

\bibitem{Bergshoeff:2004pg}
E.~Bergshoeff, A.~Collinucci, U.~Gran, D.~Roest and S.~Vandoren, \emph{{Non-extremal instantons and wormholes in string theory}}, \href{https://doi.org/10.1002/prop.200410227}{\emph{Fortsch. Phys.} {\bfseries 53} (2005) 990} [\href{https://arxiv.org/abs/hep-th/0412183}{{\ttfamily hep-th/0412183}}].

\bibitem{Andriolo:2022rxc}
S.~Andriolo, G.~Shiu, P.~Soler and T.~Van~Riet, \emph{{Axion wormholes with massive dilaton}}, \href{https://doi.org/10.1088/1361-6382/ac8fdc}{\emph{Class. Quant. Grav.} {\bfseries 39} (2022) 215014} [\href{https://arxiv.org/abs/2205.01119}{{\ttfamily 2205.01119}}].

\bibitem{Jonas:2023ipa}
C.~Jonas, G.~Lavrelashvili and J.-L.~Lehners, \emph{{Zoo of axionic wormholes}}, \href{https://doi.org/10.1103/PhysRevD.108.066012}{\emph{Phys. Rev. D} {\bfseries 108} (2023) 066012} [\href{https://arxiv.org/abs/2306.11129}{{\ttfamily 2306.11129}}].

\bibitem{Gutperle:2002km}
M.~Gutperle and W.~Sabra, \emph{{Instantons and wormholes in Minkowski and (A)dS spaces}}, \href{https://doi.org/10.1016/S0550-3213(02)00942-2}{\emph{Nucl. Phys. B} {\bfseries 647} (2002) 344} [\href{https://arxiv.org/abs/hep-th/0206153}{{\ttfamily hep-th/0206153}}].

\bibitem{Aguilar-Gutierrez:2023ril}
S.E.~Aguilar-Gutierrez, T.~Hertog, R.~Tielemans, J.P.~van~der Schaar and T.~Van~Riet, \emph{{Axion-de Sitter wormholes}}, \href{https://doi.org/10.1007/JHEP11(2023)225}{\emph{JHEP} {\bfseries 11} (2023) 225} [\href{https://arxiv.org/abs/2306.13951}{{\ttfamily 2306.13951}}].

\bibitem{Hebecker:2018ofv}
A.~Hebecker, T.~Mikhail and P.~Soler, \emph{{Euclidean wormholes, baby universes, and their impact on particle physics and cosmology}}, \href{https://doi.org/10.3389/fspas.2018.00035}{\emph{Front. Astron. Space Sci.} {\bfseries 5} (2018) 35} [\href{https://arxiv.org/abs/1807.00824}{{\ttfamily 1807.00824}}].

\bibitem{Maldacena:2004rf}
J.M.~Maldacena and L.~Maoz, \emph{{Wormholes in AdS}}, \href{https://doi.org/10.1088/1126-6708/2004/02/053}{\emph{JHEP} {\bfseries 02} (2004) 053} [\href{https://arxiv.org/abs/hep-th/0401024}{{\ttfamily hep-th/0401024}}].

\bibitem{Marolf:2022ntb}
D.~Marolf and J.E.~Santos, \emph{{The canonical ensemble reloaded: the complex-stability of Euclidean quantum gravity for black holes in a box}}, \href{https://doi.org/10.1007/JHEP08(2022)215}{\emph{JHEP} {\bfseries 08} (2022) 215} [\href{https://arxiv.org/abs/2202.11786}{{\ttfamily 2202.11786}}].

\bibitem{Liu:2023jvm}
X.~Liu, D.~Marolf and J.E.~Santos, \emph{{Stability of saddles and choices of contour in the Euclidean path integral for linearized gravity: dependence on the DeWitt parameter}}, \href{https://doi.org/10.1007/JHEP05(2024)087}{\emph{JHEP} {\bfseries 05} (2024) 087} [\href{https://arxiv.org/abs/2310.08555}{{\ttfamily 2310.08555}}].

\bibitem{Coleman:1987rm}
S.R.~Coleman, \emph{{Quantum Tunneling and Negative Eigenvalues}}, \href{https://doi.org/10.1016/0550-3213(88)90308-2}{\emph{Nucl. Phys. B} {\bfseries 298} (1988) 178}.

\bibitem{Coleman:1985rnk}
S.~Coleman, \emph{{Aspects of Symmetry}: {Selected Erice Lectures}}, Cambridge University Press, Cambridge, U.K. (1985), \href{https://doi.org/10.1017/CBO9780511565045}{10.1017/CBO9780511565045}.

\bibitem{Coleman:1977py}
S.R.~Coleman, \emph{{The Fate of the False Vacuum. 1. Semiclassical Theory}}, \href{https://doi.org/10.1103/PhysRevD.16.1248}{\emph{Phys. Rev. D} {\bfseries 15} (1977) 2929}.

\bibitem{Callan:1977pt}
C.G.~Callan, Jr. and S.R.~Coleman, \emph{{The Fate of the False Vacuum. 2. First Quantum Corrections}}, \href{https://doi.org/10.1103/PhysRevD.16.1762}{\emph{Phys. Rev. D} {\bfseries 16} (1977) 1762}.

\bibitem{Coleman:1980aw}
S.R.~Coleman and F.~De~Luccia, \emph{{Gravitational Effects on and of Vacuum Decay}}, \href{https://doi.org/10.1103/PhysRevD.21.3305}{\emph{Phys. Rev. D} {\bfseries 21} (1980) 3305}.

\bibitem{Rubakov:1996cn}
V.A.~Rubakov and O.Y.~Shvedov, \emph{{A Negative mode about Euclidean wormhole}}, \href{https://doi.org/10.1016/0370-2693(96)00766-6}{\emph{Phys. Lett. B} {\bfseries 383} (1996) 258} [\href{https://arxiv.org/abs/gr-qc/9604038}{{\ttfamily gr-qc/9604038}}].

\bibitem{Alonso:2017avz}
R.~Alonso and A.~Urbano, \emph{{Wormholes and masses for Goldstone bosons}}, \href{https://doi.org/10.1007/JHEP02(2019)136}{\emph{JHEP} {\bfseries 02} (2019) 136} [\href{https://arxiv.org/abs/1706.07415}{{\ttfamily 1706.07415}}].

\bibitem{Hertog:2018kbz}
T.~Hertog, B.~Truijen and T.~Van~Riet, \emph{{Euclidean axion wormholes have multiple negative modes}}, \href{https://doi.org/10.1103/PhysRevLett.123.081302}{\emph{Phys. Rev. Lett.} {\bfseries 123} (2019) 081302} [\href{https://arxiv.org/abs/1811.12690}{{\ttfamily 1811.12690}}].

\bibitem{Jonas:2023qle}
C.~Jonas, G.~Lavrelashvili and J.-L.~Lehners, \emph{{Stability of axion-dilaton wormholes}}, \href{https://doi.org/10.1103/PhysRevD.109.086022}{\emph{Phys. Rev. D} {\bfseries 109} (2024) 086022} [\href{https://arxiv.org/abs/2312.08971}{{\ttfamily 2312.08971}}].

\bibitem{Loges:2022nuw}
G.J.~Loges, G.~Shiu and N.~Sudhir, \emph{{Complex saddles and Euclidean wormholes in the Lorentzian path integral}}, \href{https://doi.org/10.1007/JHEP08(2022)064}{\emph{JHEP} {\bfseries 08} (2022) 064} [\href{https://arxiv.org/abs/2203.01956}{{\ttfamily 2203.01956}}].

\bibitem{Hertog:2024nys}
T.~Hertog, S.~Maenaut, B.~Missoni, R.~Tielemans and T.~Van~Riet, \emph{{Stability of axion-saxion wormholes}}, \href{https://doi.org/10.1007/JHEP11(2024)151}{\emph{JHEP} {\bfseries 11} (2024) 151} [\href{https://arxiv.org/abs/2405.02072}{{\ttfamily 2405.02072}}].

\bibitem{Liu}
X.~Liu, D.~Marolf and J.~Santos, \emph{{title TBD}},  \href{https://arxiv.org/abs/to appear}{{\ttfamily to appear}}.

\bibitem{Garriga:1997wz}
J.~Garriga, X.~Montes, M.~Sasaki and T.~Tanaka, \emph{{Canonical quantization of cosmological perturbations in the one-bubble open universe}}, \href{https://doi.org/10.1016/S0550-3213(97)00780-3}{\emph{Nucl. Phys. B} {\bfseries 513} (1998) 343} [\href{https://arxiv.org/abs/astro-ph/9706229}{{\ttfamily astro-ph/9706229}}].

\bibitem{Gratton:1999ya}
S.~Gratton and N.~Turok, \emph{{Cosmological perturbations from the no boundary Euclidean path integral}}, \href{https://doi.org/10.1103/PhysRevD.60.123507}{\emph{Phys. Rev. D} {\bfseries 60} (1999) 123507} [\href{https://arxiv.org/abs/astro-ph/9902265}{{\ttfamily astro-ph/9902265}}].

\bibitem{Kol:2006ga}
B.~Kol, \emph{{The Power of Action: The Derivation of the Black Hole Negative Mode}}, \href{https://doi.org/10.1103/PhysRevD.77.044039}{\emph{Phys. Rev. D} {\bfseries 77} (2008) 044039} [\href{https://arxiv.org/abs/hep-th/0608001}{{\ttfamily hep-th/0608001}}].

\bibitem{Horowitz}
G.T.~Horowitz, D.~Marolf and J.E.~Santos, \emph{{Constraints are not enough}},  \href{https://arxiv.org/abs/2505.13600}{{\ttfamily 2505.13600}}.

\bibitem{Weinberg:1995mt}
S.~Weinberg, \emph{{The Quantum theory of fields. Vol. 1: Foundations}}, Cambridge University Press (6, 2005), \href{https://doi.org/10.1017/CBO9781139644167}{10.1017/CBO9781139644167}.

\bibitem{VanRiet:2020pcn}
T.~Van~Riet, \emph{{Instantons, Euclidean wormholes and AdS/CFT}}, \href{https://doi.org/10.22323/1.376.0121}{\emph{PoS} {\bfseries CORFU2019} (2020) 121} [\href{https://arxiv.org/abs/2004.08956}{{\ttfamily 2004.08956}}].

\bibitem{Burgess:1989da}
C.P.~Burgess and A.~Kshirsagar, \emph{{Wormholes and Duality}}, \href{https://doi.org/10.1016/0550-3213(89)90186-7}{\emph{Nucl. Phys. B} {\bfseries 324} (1989) 157}.

\bibitem{Andrade:2011dg}
T.~Andrade and D.~Marolf, \emph{{AdS/CFT beyond the unitarity bound}}, \href{https://doi.org/10.1007/JHEP01(2012)049}{\emph{JHEP} {\bfseries 01} (2012) 049} [\href{https://arxiv.org/abs/1105.6337}{{\ttfamily 1105.6337}}].

\bibitem{Bergshoeff:2004fq}
E.~Bergshoeff, A.~Collinucci, U.~Gran, D.~Roest and S.~Vandoren, \emph{{Non-extremal D-instantons}}, \href{https://doi.org/10.1088/1126-6708/2004/10/031}{\emph{JHEP} {\bfseries 10} (2004) 031} [\href{https://arxiv.org/abs/hep-th/0406038}{{\ttfamily hep-th/0406038}}].

\bibitem{xPand}
C.~Pitrou, X.~Roy and O.~Umeh, \emph{{xPand: An algorithm for perturbing homogeneous cosmologies}}, \href{https://doi.org/10.1088/0264-9381/30/16/165002}{\emph{Class. Quant. Grav.} {\bfseries 30} (2013) 165002} [\href{https://arxiv.org/abs/1302.6174}{{\ttfamily 1302.6174}}].

\bibitem{xtensor}
J.M.~Mart{\'i}n-Garc{\'i}a, \emph{{xTensor: A free fast abstract tensor manipulator}},  in \emph{The Eleventh Marcel Grossmann Meeting: On Recent Developments in Theoretical and Experimental General Relativity, Gravitation and Relativistic Field Theories (In 3 Volumes)}, pp.~1552--1554, World Scientific, 2008.

\bibitem{Brizuela:2008ra}
D.~Brizuela, J.M.~Martin-Garcia and G.A.~Mena~Marugan, \emph{{xPert: Computer algebra for metric perturbation theory}}, \href{https://doi.org/10.1007/s10714-009-0773-2}{\emph{Gen. Rel. Grav.} {\bfseries 41} (2009) 2415} [\href{https://arxiv.org/abs/0807.0824}{{\ttfamily 0807.0824}}].

\bibitem{Maenaut:2020Git}
S.~Maenaut, ``{xPand} modifications for {Euclidean} spacetimes.'' \url{https://github.com/SimonMaenaut/xPand}.

\bibitem{Gratton:2000fj}
S.~Gratton and N.~Turok, \emph{{Homogeneous modes of cosmological instantons}}, \href{https://doi.org/10.1103/PhysRevD.63.123514}{\emph{Phys. Rev. D} {\bfseries 63} (2001) 123514} [\href{https://arxiv.org/abs/hep-th/0008235}{{\ttfamily hep-th/0008235}}].

\bibitem{Gilbert}
G.T.~Gilbert, \emph{{Positive Definite Matrices and Sylvester's Criterion}}, {\emph{The American Mathematical Monthly} {\bfseries 98} (1991) 44}.

\end{thebibliography}\endgroup

\end{document}